\newlength{\absize}
\documentclass[12pt]{article}
\usepackage{latexsym}
\usepackage{amssymb}
\usepackage{epsf}
\usepackage{setspace}
\renewcommand{\arraystretch}{2.5}

\newcommand{\figsize}{\small}

\input prepictex
\input pictex
\input postpictex
\newdimen\tdim
\tdim=\unitlength
\def\stpltsmbl{\setplotsymbol ({\small .})}

\newbox\sru
\setbox\sru=\hbox{\beginpicture
\setcoordinatesystem units <\tdim,\tdim>
\stpltsmbl
\setquadratic
\plot
  0.0   0.0
  4.8   1.5
  7.5   5.0
  7.3   8.5
  5.0  10.0
  2.7   8.5
  2.5   5.0
  5.2   1.5
 10.0   0.0
/
\endpicture}
\def\springru #1 #2 *#3 /{\multiput {\copy\sru}  at
#1 #2 *#3 10 0 /}

\newcommand{\real}{{\rm Re\,}}
\renewcommand{\bar}{\overline}

\newcommand{\un}{\mathcal{U}}
\newcommand{\bz}{\mathcal{BZ}}

\def\ltilde#1{\mathord{\mathop{\kern 0pt #1}\limits_{\sim\atop}}}

\begin{document}

\thispagestyle{empty}
\pagestyle{empty}
\renewcommand{\thefootnote}{\fnsymbol{footnote}}
\newcommand{\starttext}{\newpage\normalsize
 \pagestyle{plain}
 \setlength{\baselineskip}{3ex}\par
 \setcounter{footnote}{0}
 \renewcommand{\thefootnote}{\arabic{footnote}}
 }
\newcommand{\preprint}[1]{\begin{flushright}
 \setlength{\baselineskip}{3ex}#1\end{flushright}}
\renewcommand{\title}[1]{\begin{center}\LARGE
 #1\end{center}\par}
\renewcommand{\author}[1]{\vspace{2ex}{\Large\begin{center}
 \setlength{\baselineskip}{3ex}#1\par\end{center}}}
\renewcommand{\thanks}[1]{\footnote{#1}}
\renewcommand{\abstract}[1]{\vspace{2ex}\normalsize\begin{center}
 \centerline{\bf Abstract}\par\vspace{2ex}\parbox{\absize}{#1
 \setlength{\baselineskip}{2.5ex}\par}
 \end{center}}

\preprint{}
\title{Another Odd Thing About Unparticle Physics}
\author{
Howard~Georgi\thanks{\noindent \tt georgi@physics.harvard.edu}
 \\ \medskip
Center for the Fundamental Laws of Nature\\
Jefferson Physical Laboratory \\
Harvard University \\
Cambridge, MA 02138\\
~\\ 
April 2007}
\abstract{The peculiar propagator of scale invariant unparticles has phases
that produce unusual patterns of interference with standard model
processes. We illustrate some of these effects in $e^+e^-\to\mu^+\mu^-$.}

\starttext

\setcounter{equation}{0}
\section{\label{intro}Introduction}

In a previous paper~\cite{Georgi:2007ek}, I argued that a scale invariant
sector that decouples at a large scale is associated with ``unparticles''
whose production might be detectable in missing
energy and momentum distributions. In this note, we consider 
some of the leading
virtual effects of unparticles. In particular, we write down
the unparticle
propagator and consider the interference between standard model amplitudes
and amplitudes involving virtual unparticles. 
As emphasized long ago by Eichten, Lane and Peskin~\cite{Eichten:1983hw}, 
this kind of interference can be
a sensitive probe of high-energy processes. In particular, the interference
terms are effects of leading nontrivial order in the small couplings of
unparticles to standard model particles, the same order as the production
cross-sections considered in \cite{Georgi:2007ek}.
We will find that the peculiar phases
associated with the unparticle propagator 
in the time-like region give rise to unusual patterns
of interference which depend dramatically on the scaling dimension of the
unparticles. 

In this note, as in~\cite{Georgi:2007ek}, we assume
that the very high energy
theory contains the fields
of the standard model and the fields of a theory with a
nontrivial IR fixed point, which I call $\bz$ (for
Banks-Zaks~\cite{Banks:1981nn}) fields. 
The two sets interact through the exchange of particles with a 
large mass scale $M_{\un}$. Below $M_{\un}$, there are
nonrenormalizable couplings between standard model
fields and Banks-Zaks
fields suppressed by powers of $M_{\un}$.
The renormalizable couplings of the $\bz$ fields then produce
dimensional transmutation and the scale-invariant unparticle
fields emerge below an energy
scale $\Lambda_{\un}$.
In the effective theory below the scale $\Lambda_{\un}$ the $\bz$ operators
match onto unparticle operators, and the nonrenormalizable interactions
match onto a new set of interactions between standard model and unparticle
fields with small coefficients. 
We make crucial use of the
simplifications that result by working to lowest nontrivial order in the
small couplings of unparticles fields to standard model fields in the
effective field theory below $\Lambda_{\un}$. This allows us reliably to
calculate some important quantities without having to understand in detail
what unparticles look like. We will return to some of these questions at
the end of the paper.

To illustrate the interesting properties of the unparticle propagator,
we consider the example of the low energy effect of the following 
interaction terms.
\begin{equation}
\frac{C_{V\un}\,\Lambda_{\un}^{k+1-d_{\un}}}{M_{\un}^k}\;
\bar e\,\gamma_\mu\,e\,O^\mu_{\un}
+
\frac{C_{A\un}\,\Lambda_{\un}^{k+1-d_{\un}}}{M_{\un}^k}\;
\bar e\,\gamma_\mu\gamma_5\,e\,O^\mu_{\un}
\label{eandu}
\end{equation}
where the unparticle operator is hermitian and transverse,
\begin{equation}
\partial_\mu O^\mu_{\un}=0\,.
\label{transverse}
\end{equation}
I stress that this is just an example. Unparticle operators with different
tensor structures can be dealt with in a similar way.

In the notation of \cite{Georgi:2007ek}, the transverse 4-vector
unparticle propagator is given by
\begin{equation}
\begin{array}{c}
\displaystyle
\int\,e^{iPx}\,
\left\langle0\right|T(O^\mu_{\un}(x)\,O^\nu_{\un}(0))\left|0\right\rangle\,d^4x
\\ \displaystyle
=
i\frac{A_{d_{\un}}}{2\pi}\,
\int_0^\infty\,
\left(M^2\right)^{d_{\un}-2}\,
\frac{-g^{\mu\nu}+P^\mu P^\nu/P^2}{P^2-M^2+i\epsilon}\,
dM^2
\\ \displaystyle
=
i\frac{A_{d_{\un}}}{2}\,
\frac{-g^{\mu\nu}+P^\mu P^\nu/P^2}{\sin(d_{\un}\pi)}\,
\left(-P^2-i\epsilon\right)^{d_{\un}-2}
\end{array}
\label{propagator}
\end{equation}
where
\begin{equation}
A_{d_{\un}}=\frac{16\pi^{5/2}}{(2\pi)^{2d_{\un}}}
\,\frac{\Gamma(d_{\un}+1/2)}{\Gamma(d_{\un}-1)\,\Gamma(2d_{\un})}
\label{adu}
\end{equation}
We can check this odd-looking result by finding the
discontinuity across the cut for $P^2>0$.
\begin{equation}
\begin{array}{c}
\displaystyle
i\frac{A_{d_{\un}}}{2}\,\frac{-g^{\mu\nu}+P^\mu P^\nu/P^2}{\sin(d_{\un}\pi)}\,
\left(P^2\right)^{d_{\un}-2}
\left(
\left(-1-i\epsilon\right)^{d_{\un}-2}
-\left(-1+i\epsilon\right)^{d_{\un}-2}
\right)
\\ \displaystyle
=i\frac{A_{d_{\un}}}{2}\,
\frac{-g^{\mu\nu}+P^\mu P^\nu/P^2}{\sin(d_{\un}\pi)}\,
\left(P^2\right)^{d_{\un}-2}
\left(
e^{-i(d_{\un}-2)\pi}
-e^{i(d_{\un}-2)\pi}
\right)
\\ \displaystyle
i\frac{A_{d_{\un}}}{2}\,\frac{-g^{\mu\nu}+P^\mu P^\nu/P^2}{\sin(d_{\un}\pi)}\,
\left(P^2\right)^{d_{\un}-2}
\left(-2i\sin(d_{\un}\pi)\right)
=A_{d_{\un}}\,\Bigl(-g^{\mu\nu}+P^\mu P^\nu/P^2\Bigr)\,
\left(P^2\right)^{d_{\un}-2}
\end{array}
\label{discontinuity}
\end{equation}
in agreement with the arguments of \cite{Georgi:2007ek}. 

As
(\ref{discontinuity}) shows, the non-trivial phases along the physical cut
in (\ref{propagator})
play a necessary role in reproducing the scale invariance.
We will find that 
these phases, even more than the nontrivial scaling itself, produce unique
physical effects in interference. These peculiar interference effects are
the key results in this paper.
We explore a few of these below for the explicit
example of (\ref{eandu}).

It is important to note that while the discontinuity across the cut is not
singular for integer $d_{\un}>1$, the propagator (\ref{propagator}) is
singular because of the $\sin(d_{\un}\pi)$ in the denominator. I believe
that this is a real effect. These integer values describe multiparticle
cuts and the mathematics is telling us that we should
not be trying to describe them with a single unparticle field. For this
reason we will focus on $1<d_{\un}<2$, and we will find that the virtual
effect of unparticles are gentlest away from the integer boundaries.

Let us first compute the cross section for $e^+e^-\to\mu^+\mu^-$ in the
the presence of the interactions (\ref{eandu}). It is convenient to rescale
the dimensional coefficients to the $Z$ mass, and define the dimensionless
coefficients 
\begin{equation}
{c_{V\un}}=
\frac{C_{V\un}\,\Lambda_{\un}^{k+1-d_{\un}}}{M_{\un}^k\,M_Z^{1-d_{\un}}}
\quad\quad
{c_{A\un}}=
\frac{C_{A\un}\,\Lambda_{\un}^{k+1-d_{\un}}}{M_{\un}^k\,M_Z^{1-d_{\un}}}
\label{cs}
\end{equation}

Then (ignoring the lepton masses) the square of the invariant matrix element
can be written as (where
${q^2}={s}$ is the square of the total center of mass
energy and $\theta$ is the angle of the $\mu^-$ from the $e^-$ direction in
the center of mass)
\begin{equation}
{\renewcommand{\arraystretch}{2.5}
\begin{array}{c}
\displaystyle
|\mathcal{M}|^2
=2(q^2)^2\left[
\Bigl(
\left|\Delta_{VV}(q^2)\right|^2
+\left|\Delta_{AA}(q^2)\right|^2
+\left|\Delta_{VA}(q^2)\right|^2
+\left|\Delta_{AV}(q^2)\right|^2
\Bigr)\,(1+\cos^2\theta)
\right.
\\ \displaystyle
+\left.
\Bigl(
\real(\Delta^*_{VV}(q^2)\,\Delta_{AA}(q^2))
+\real(\Delta^*_{VA}(q^2)\,\Delta_{AV}(q^2))
\Bigr)\,4\cos\theta
\right]
\end{array}}
\label{me}
\end{equation}
where
\begin{equation}
\Delta_{xy}(q^2)\equiv \sum_{j=\gamma\atop Z,\un}
d^e_{xj}{d^\mu_{yj}}^*\,
\Delta_j(q^2)\quad\mbox{where $x,y=V$ or $A$}
\label{amp}
\end{equation}
with the $d$s given in the following table
\begin{equation}
{\renewcommand{\arraystretch}{2}
\begin{array}{|c||c|c|c|}
\hline
d_{xj}&\gamma&Z&\un\\
\hline
\hline
\rule[-3ex]{0pt}{1ex}
V&e&\displaystyle\frac{e}{\sin\theta\cos\theta}(-1/4+\sin^2\theta)
&\displaystyle\frac{c_{V\un}}{M_Z^{d_{\un}-1}}\\
\hline
\rule[-3ex]{0pt}{1ex}
A&0&\displaystyle\frac{e/4}{\sin\theta\cos\theta}
&\displaystyle\frac{c_{A\un}}{M_Z^{d_{\un}-1}}\\
\hline
\end{array}}
\label{ds}
\end{equation}
and the $\Delta_j$s being the $\gamma$, $Z$ and $\un$ propagators,
\begin{equation}
{\renewcommand{\arraystretch}{2}
\begin{array}{|c||c|c|c|}
\hline
\Delta_{j}&\gamma&Z&\un\\
\hline
\hline
\rule[-3ex]{0pt}{1ex}
=&\displaystyle\frac{1}{q^2}
&\displaystyle\frac{1}{(q^2-M_Z^2+iM_Z\Gamma_Z)}
&\displaystyle\frac{A_{d_{\un}}}{2\sin(d_{\un}\pi)}\,
\left(q^2\right)^{d_{\un}-2}\,e^{-i(d_\un-2)\pi}\\
\hline
\hline
\end{array}}
\end{equation}
{\figsize\begin{figure}[htb]
$$\beginpicture
\setcoordinatesystem units <.8\tdim,.8\tdim>
\put {{\epsfxsize=320\tdim \epsfbox{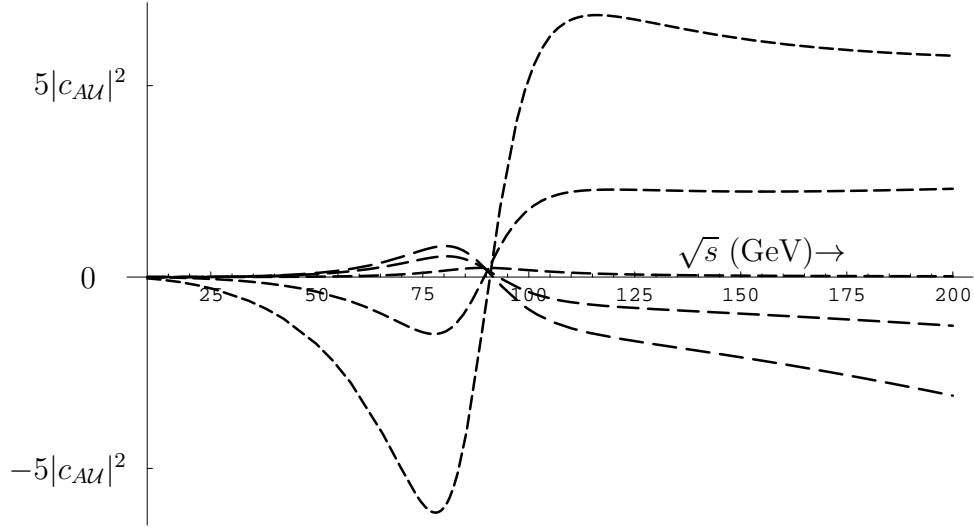}}} at 0 0
\put {$5|c_{A\un}|^2$} [r] at -200 84
\put {$-5|c_{A\un}|^2$} [r] at -200 -98
\put {$0$~~~} [r] at -200 -7
\put {$\sqrt{s}$~(GeV)$\to$} at 100 5
\endpicture$$
\caption{\figsize\sf\label{fig-1}The fractional change in total
cross-section for $e^+e^-\to\mu^+\mu^-$ versus $\sqrt{s}$ for $d_{\un}=1.1$,
$1.3$, $1.5$, $1.7$ and $1.9$
for non-zero $c_{A\un}$ and $c_{V\un}=0$. 
The dash-length increases with $d_{\un}$.}
\end{figure}}
We have tacitly assumed in (\ref{ds}) that the unparticle interactions are
lepton-flavor-blind, so that we do not have to keep track of the $e$ and
$\mu$ superscripts on the $c$s, and we will continue to assume this in the
graphs below. But (\ref{me}) is entirely general and does not depend on
this assumption.

As a first example of the interesting structure of (\ref{me}), consider the
total cross section in the LEP region. We are used to thinking that the $Z$
pole is not a good place to look for interference with the effects of
small non-renormalizable interactions because the amplitude is dominantly
imaginary on the pole. This prejudice is not warranted for unparticle
interactions. The unparticle amplitude can interfere with both the real and
imaginary parts of the standard model and can therefore contribute both on
and off the pole.

It is instructive to 
begin by assuming $c_{V\un}=0$ (remember, we are taking the same $c$ for
$e$ and $\mu$) and considering the total cross section. 
Because the vector coupling vanishes, the 
interference between the unparticle exchange amplitude
and the photon decay amplitude does not contribute to the total cross
section, so we expect only interference with $Z$ exchange.
In figure~\ref{fig-1}, I show the fractional change in the
total cross section for small non-zero $c_{A\un}$ for various values of
$d_{\un}$ between $1$ and $2$. The dominant effect as
expected is the interference term proportional to a single power of
$|c_{A\un}|^2$. But the striking thing about this graph is how sensitively
the result depends on the value of $d_{\un}$. We can understand qualitatively
what is going on by thinking about
the phase of the unparticle propagator along the physical cut which is
\begin{equation}
\phi_{d_\un}=-(d_{\un}-1)\pi
\label{phase}
\end{equation}
{\figsize\begin{figure}[htb]
$$\beginpicture
\setcoordinatesystem units <.8\tdim,.8\tdim>
\put {{\epsfxsize=320\tdim \epsfbox{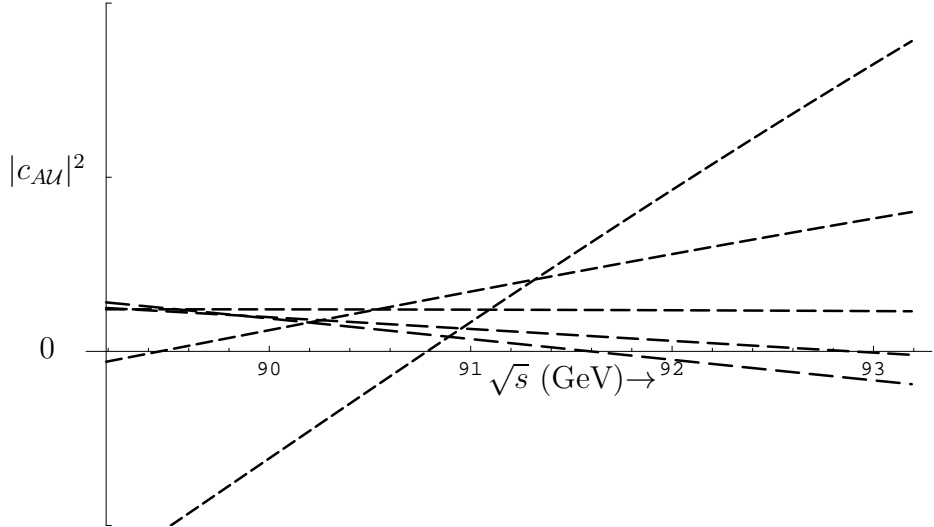}}} at 0 0
\put {$|c_{A\un}|^2$} [r] at -200 44
\put {$0$~~~} [r] at -200 -40
\put {$\sqrt{s}$~(GeV)$\to$} at 30 -55
\endpicture$$
\caption{\figsize\sf\label{fig-1-2}The fractional change in total
cross-section for $e^+e^-\to\mu^+\mu^-$ versus $\sqrt{s}$ for $d_{\un}=1.1$,
$1.3$, $1.5$, $1.7$ and $1.9$
for non-zero $c_{A\un}$ and $c_{V\un}=0$. The dash-length increases with
$d_{\un}$. Note the different scales compared to figure~\protect\ref{fig-1}.}
\end{figure}}

The real part of (\ref{phase}) is positive for $1<d_\un<3/2$ and negative
for $3/2<d_\un<2$. The real part of $1/(q^2-M_Z^2+iM_Z\Gamma_Z)$ is
negative below the $Z$ pole and positive above. Thus away from the $Z$
pole, where the imaginary part of $1/(q^2-M_Z^2+iM_Z\Gamma_Z)$ is small, we
expect destructive (constructive) interference below (above) the pole for
$1<d_\un<3/2$, and vice-versa for $3/2<d_\un<1$.
Near the $Z$ pole, the situation is complicated, as illustrated in
figure~\ref{fig-1-2} because both real and 
imaginary parts contribute to the interference.
{\figsize\begin{figure}[htb]
$$\beginpicture
\setcoordinatesystem units <.8\tdim,.8\tdim>
\put {{\epsfxsize=320\tdim \epsfbox{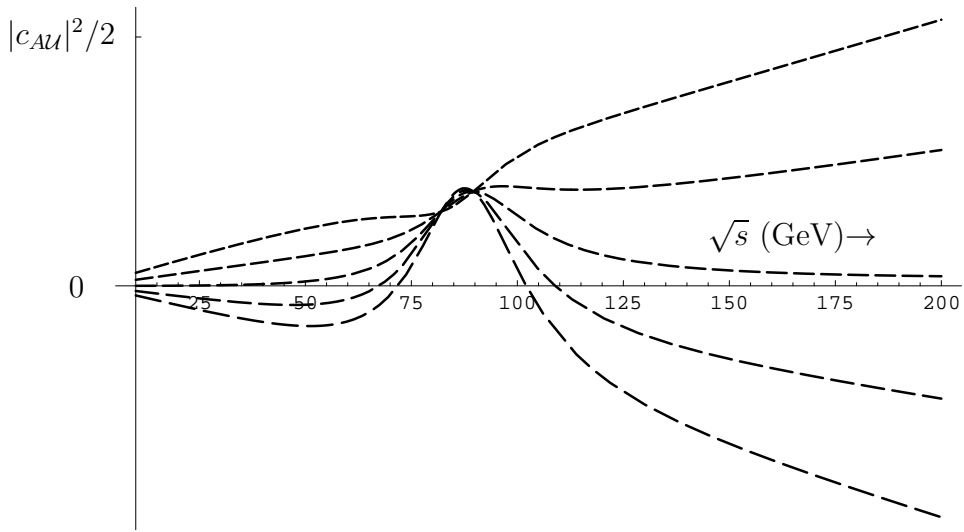}}} at 0 0
\put {$|c_{A\un}|^2/2$} [r] at -200 110
\put {$0$~~~} [r] at -200 -9
\put {$\sqrt{s}$~(GeV)$\to$} at 120 15
\endpicture$$
\caption{\figsize\sf\label{fig-2}The fractional change in total
cross-section for $e^+e^-\to\mu^+\mu^-$ versus $\sqrt{s}$ for $d_{\un}=1.48$,
$1.49$, $1.5$, $1.51$ and $1.52$
for non-zero $c_{A\un}$ and $c_{V\un}=0$. 
The dash-length increases with $d_{\un}$.
Note the different vertical scale compared to figure~\protect\ref{fig-1}.}
\end{figure}}

The situation simplifies in a very interesting way for $d_{\un}=3/2$. In
this case, the phase from (\ref{phase}) is $\phi_{d_\un}=-\pi/2$, so the
unparticle amplitude interferes only with the imaginary part of the
$Z$-exchange amplitude. This is a smaller effect than we see for values of
$d_{\un}$ very different from $3/2$ because it is proportional
to the $Z$ width, rather than $q^2-M_Z^2$. 
It gives constructive interference that 
peaks on the $Z$ pole and goes to zero far from the pole. This is shown on
a different scale in figure~\ref{fig-2}. Here I have also included a few
values of $d_{\un}$ close to $3/2$, for comparison.

{\figsize\begin{figure}[htb]
$$\beginpicture
\setcoordinatesystem units <.8\tdim,.8\tdim>
\put {{\epsfxsize=320\tdim \epsfbox{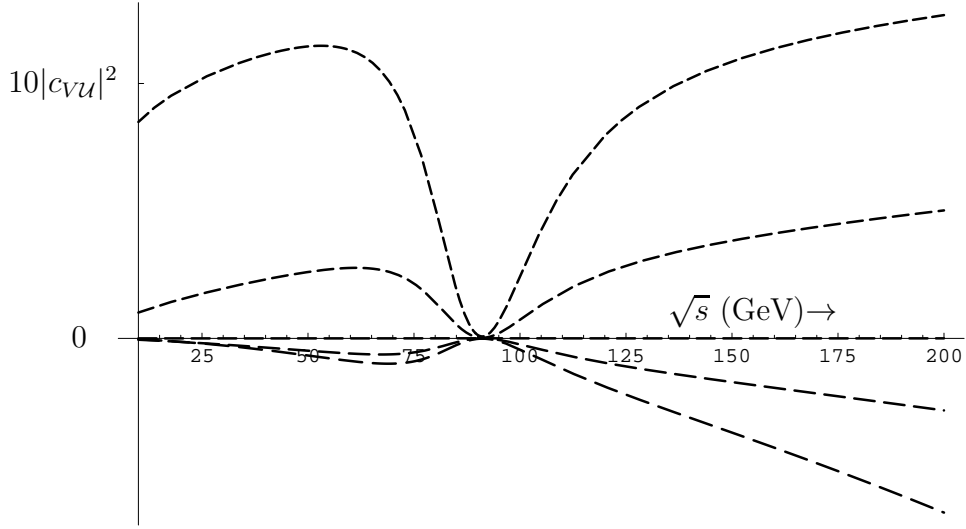}}} at 0 0
\put {$10|c_{V\un}|^2$} [r] at -200 85
\put {$0$~~~} [r] at -200 -35
\put {$\sqrt{s}$~(GeV)$\to$} at 100 -23
\endpicture$$
\caption{\figsize\sf\label{fig-3}The fractional change in total
cross-section for $e^+e^-\to\mu^+\mu^-$ versus $\sqrt{s}$ for $d_{\un}=1.1$,
$1.3$, $1.5$, $1.7$ and $1.9$
for non-zero $c_{V\un}$ and $c_{A\un}=0$. The dash-length increases with
$d_{\un}$.}
\end{figure}}
Having seen how things work for purely axial vector unparticle couplings,
let us now consider what the total cross section looks like for a vector
coupling. Now we expect interference with the photon-exchange amplitude,
and because the vector part of the leptonic coupling of the $Z$ is (by an
``accident'' of the value of $\sin^2\theta$) very small, there is very
little interference with the $Z$-exchange amplitude. Now we expect
constructive interference for 
$1<d_\un<3/2$ and destructive interference for $3/2<d_\un<1$.
The result is shown in figure~\ref{fig-3}. The dip at the $Z$ pole arises
simply because we are plotting a fractional change and the large contribution
from the pole is in the denominator.

{\figsize\begin{figure}[htb]
$$\beginpicture
\setcoordinatesystem units <.8\tdim,.8\tdim>
\put {{\epsfxsize=320\tdim \epsfbox{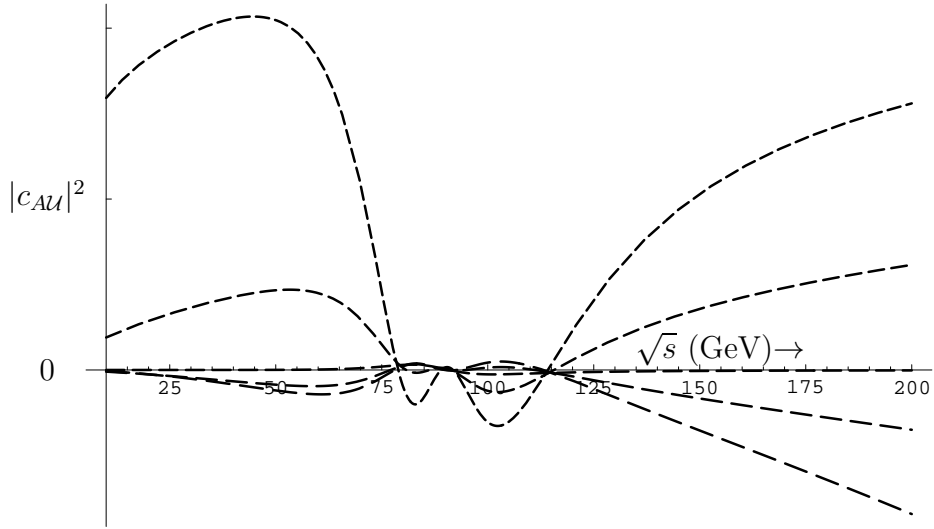}}} at 0 0
\put {$|c_{A\un}|^2$} [r] at -200 32
\put {$0$~~~} [r] at -200 -50
\put {$\sqrt{s}$~(GeV)$\to$} at 100 -40
\endpicture$$
\caption{\figsize\sf\label{fig-fba}The change in the front-back
asymmetry for $e^+e^-\to\mu^+\mu^-$ versus $\sqrt{s}$ for $d_{\un}=1.1$,
$1.3$, $1.5$, $1.7$ and $1.9$
for non-zero $c_{A\un}$ and $c_{V\un}=0$. The dash-length increases with
$d_{\un}$.}
\end{figure}}
{\figsize\begin{figure}[htb]
$$\beginpicture
\setcoordinatesystem units <.8\tdim,.8\tdim>
\put {{\epsfxsize=320\tdim \epsfbox{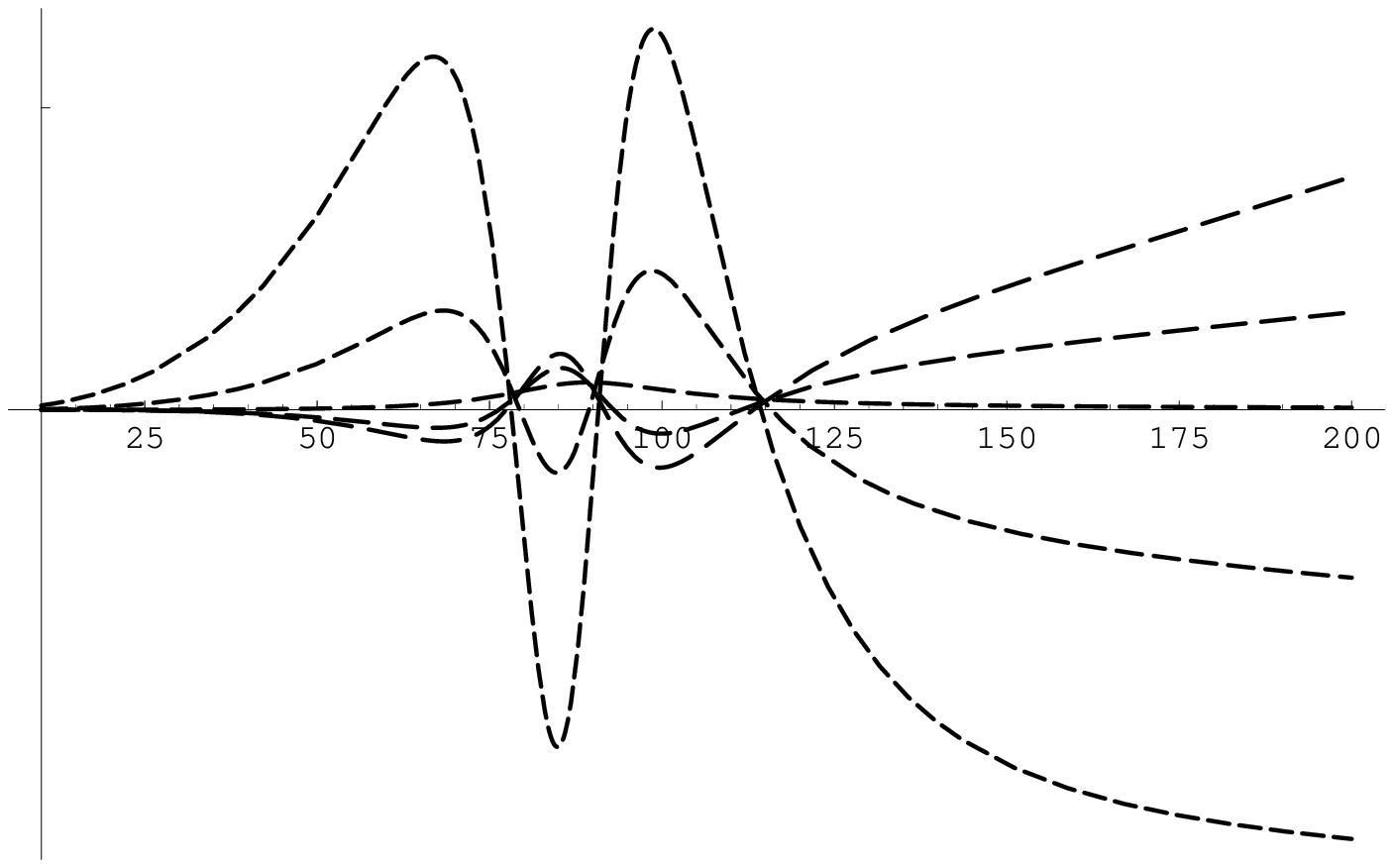}}} at 0 0
\put {$|c_{V\un}|^2/2$} [r] at -200 95
\put {$0$~~~} [r] at -200 5
\put {$\sqrt{s}$~(GeV)$\to$} at 100 -13
\endpicture$$
\caption{\figsize\sf\label{fig-fbv}The change in the front-back
asymmetry for $e^+e^-\to\mu^+\mu^-$ versus $\sqrt{s}$ for $d_{\un}=1.1$,
$1.3$, $1.5$, $1.7$ and $1.9$
for non-zero $c_{V\un}$ and $c_{A\un}=0$. The dash-length increases with
$d_{\un}$.}
\end{figure}}
The unparticle interference in the
matrix element (\ref{me}) also gives rise to a complicated pattern of
changes in the front-back asymmetry
\begin{equation}
\frac{\sigma_f-\sigma_b}{\sigma_f+\sigma_b}
=\frac{3}{8}
\left(\frac{\real(\Delta^*_{VV}(q^2)\,\Delta_{AA}(q^2))
+\real(\Delta^*_{VA}(q^2)\,\Delta_{AV}(q^2))
}{\left|\Delta_{VV}(q^2)\right|^2
+\left|\Delta_{AA}(q^2)\right|^2
+\left|\Delta_{VA}(q^2)\right|^2
+\left|\Delta_{AV}(q^2)\right|^2}\right)
\label{fba}
\end{equation}
This is shown in figures~\ref{fig-fba}
and \ref{fig-fbv}. As for the total cross section, the effect for
$d_{\un}=3/2$ is smaller and concentrated at the $Z$ pole. 
In figures~\ref{fig-fba}
and \ref{fig-fbv}, we focus down on values of $d_{\un}\approx3/2$.

{\figsize\begin{figure}[htb]
$$\beginpicture
\setcoordinatesystem units <.8\tdim,.8\tdim>
\put {{\epsfxsize=320\tdim \epsfbox{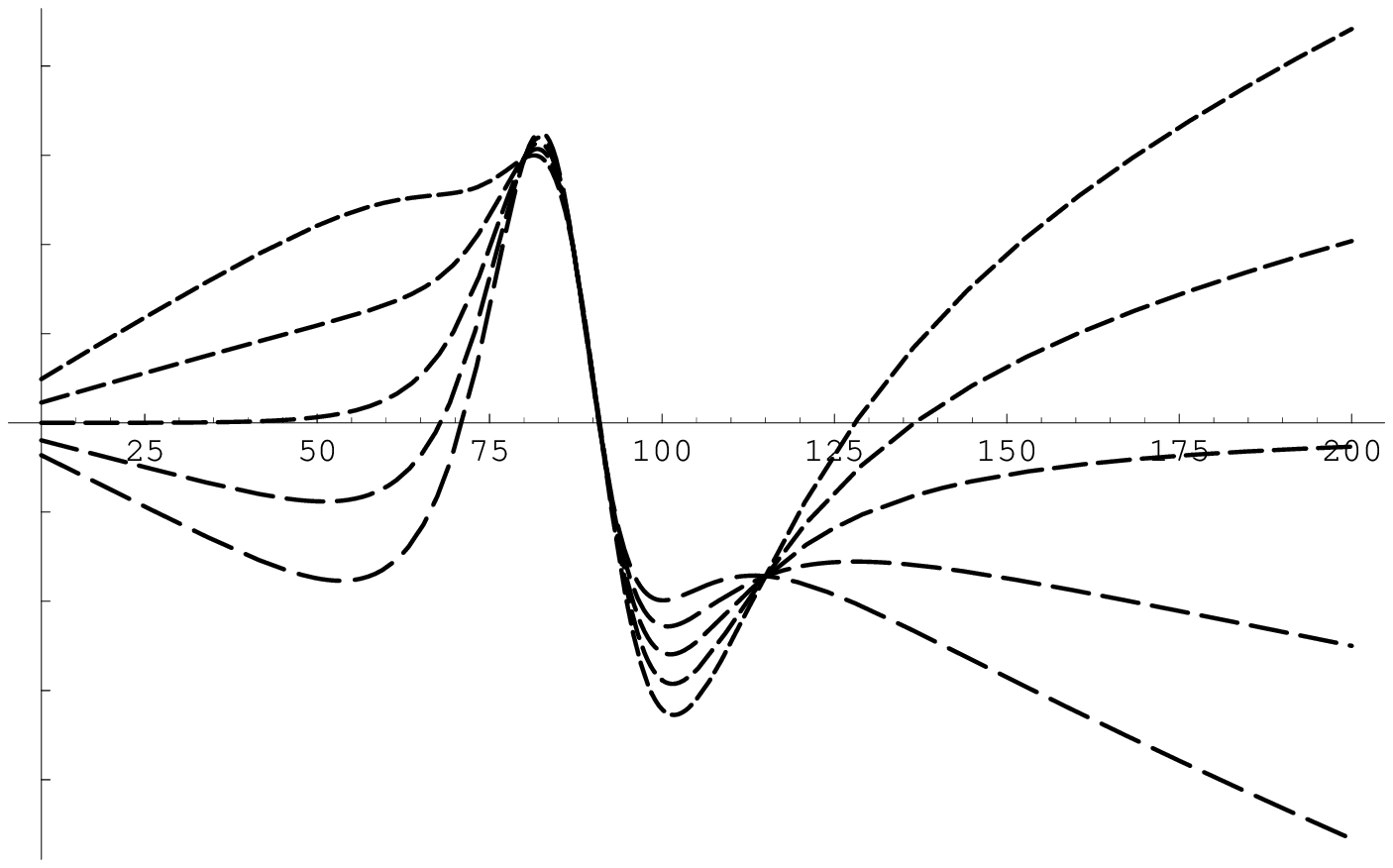}}} at 0 0
\put {$|c_{A\un}|^2/100$} [r] at -200 29
\put {$0$~~~} [r] at -200 3
\put {$\sqrt{s}$~(GeV)$\to$} at 12 17
\endpicture$$
\caption{\figsize\sf\label{fig-fba5}The change in the front-back
asymmetry for $e^+e^-\to\mu^+\mu^-$ versus $\sqrt{s}$ for $d_{\un}=1.48$,
$1.49$, $1.5$, $1.51$ and $1.52$
for non-zero $c_{A\un}$ and $c_{V\un}=0$. The dash-length increases with
$d_{\un}$.}
\end{figure}}
{\figsize\begin{figure}[htb]
$$\beginpicture
\setcoordinatesystem units <.8\tdim,.8\tdim>
\put {{\epsfxsize=320\tdim \epsfbox{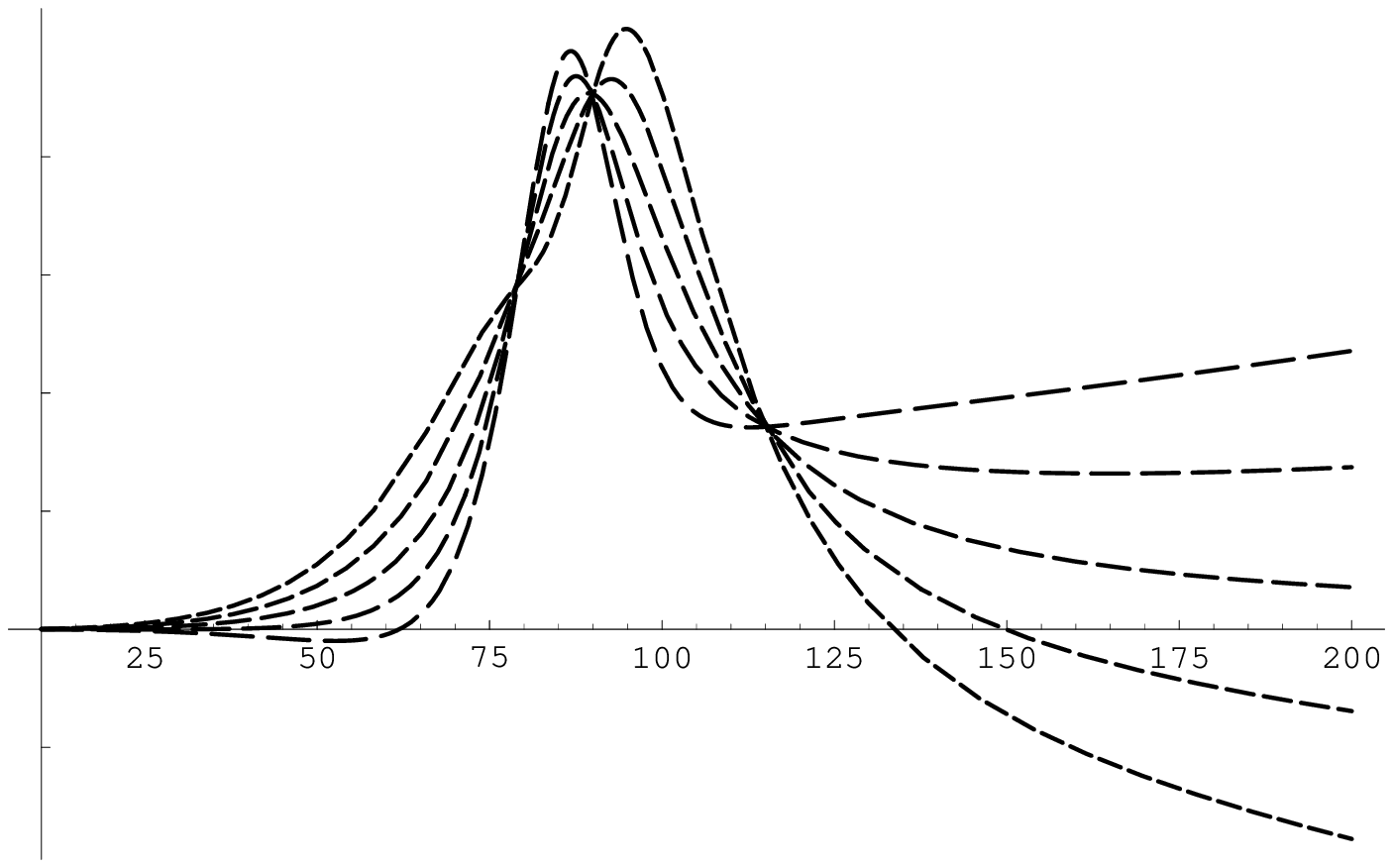}}} at 0 0
\put {$|c_{V\un}|^2/100$} [r] at -200 -23
\put {$0$~~~} [r] at -200 -57
\put {$\sqrt{s}$~(GeV)$\to$} at -15 -45
\endpicture$$
\caption{\figsize\sf\label{fig-fbv5}The change in the front-back
asymmetry for $e^+e^-\to\mu^+\mu^-$ versus $\sqrt{s}$ for $d_{\un}=1.48$,
$1.49$, $1.5$, $1.51$ and $1.52$
for non-zero $c_{V\un}$ and $c_{A\un}=0$. The dash-length increases with
$d_{\un}$.}
\end{figure}}

I hope I have convinced the reader that the unparticle propagator in the
time-like region has interesting properties that force us to reexamine
many of our preconceived notions about interference. 
Working to lowest non-trivial order in the couplings of the unparticles in
the effective low energy theory, we can make detailed predictions of the
form of interference between time-like unparticle exchange amplitudes and
standard model amplitudes even though we still lack an intuitive or even
detailed picture of what an unparticle looks like.

Let me close with a couple of more
speculative comments.
One might argue that the term
``propagator'' is not particularly felicitous for 
the unparticle time-ordered product, (\ref{propagator}),
because the unparticle does not really propagate in the usual way. 
It is also worth noting
the connection between this analysis and the more confusing issue of
unparticle decay. There is a sense in which the 
unparticle exchange amplitude that we have used in
our analysis is associated with unparticle production and decay. But in
the process we have studied in this note, 
the decay process is masked by the leading order (and therefore larger) 
interference term. And as with
the term ``propagator,'' the term
``decay'' may be a little misleading for an unparticle because it suggests
that the particle was propagating over an large distance before it
decayed. I hope to return to these deliciously confusing issues in a future
publication.

\section*{Acknowledgments}

I am grateful to Nima Arkani-Hamed,
Spencer Chang, Lisa Randall and Matthew Schwartz for discussions.
An independent work addressing 
some of the same issue is \cite{Cheung:2007ue}.

This research is supported in part by
the National Science Foundation under grant PHY-0244821.


\begin{thebibliography}{1}

\bibitem{Georgi:2007ek}
H.~Georgi, ``Unparticle physics,''
\href{http://www.arXiv.org/abs/hep-ph/0703260}{{\tt hep-ph/0703260}}.

\bibitem{Eichten:1983hw}
E.~Eichten, K.~D. Lane, and M.~E. Peskin, ``New tests for quark and lepton
  substructure,'' {\em Phys. Rev. Lett.} {\bf 50} (1983)
811--814.

\bibitem{Banks:1981nn}
T.~Banks and A.~Zaks, ``On the phase structure of vector - like gauge theories
  with massless fermions,'' {\em Nucl. Phys.} {\bf B196} (1982)
189.

\bibitem{Cheung:2007ue}
K.~Cheung, W.-Y. Keung, and T.-C. Yuan, ``Novel signals in unparticle
  physics,''
\href{http://www.arXiv.org/abs/arXiv:0704.2588 [hep-ph]}{{\tt arXiv:0704.2588
  [hep-ph]}}.

\end{thebibliography}


\providecommand{\href}[2]{#2}\begingroup\raggedright\endgroup

\end{document}